\pdfoutput=1
\documentclass[pdflatex,sn-mathphys-num]{sn-jnl}
\usepackage{graphicx}
\usepackage{multirow}
\usepackage{amsmath,amssymb,amsfonts}
\usepackage{amsthm}
\usepackage{mathrsfs}
\usepackage[title]{appendix}
\usepackage{xcolor}
\usepackage{textcomp}
\usepackage{manyfoot}
\usepackage{booktabs}
\usepackage{algorithm}
\usepackage{algorithmicx}
\usepackage{algpseudocode}
\usepackage{listings}
\usepackage{lmodern}%

\unnumbered

\begin{document}

\title[Phlystron – A photonic terahertz amplifier]{Phlystron – A photonic terahertz amplifier}

\author[1,2]{\fnm{Christian} \sur{Rentschler}}
\author[1]{\fnm{Nicholas H.} \sur{Matlis}}
\author[1,3]{\fnm{Umit} \sur{Demirbas}}
\author[1]{\fnm{Zhelin} \sur{Zhang}}
\author[1,2]{\fnm{Jonas} \sur{Nitzsche}}
\author[1]{\fnm{Koustuban} \sur{Ravi}}
\author[1]{\fnm{Mikhail} \sur{Pergament}}
\author*[1,2,4]{\fnm{Franz X.} \sur{Kärtner}}\email{franz.kaertner@desy.de}

\affil[1]{\orgdiv{Center for Free-Electron Laser Science CFEL}, \orgname{Deutsches Elektronen-Synchrotron DESY}, \orgaddress{\street{Notkestrasse 85}, \city{Hamburg}, \postcode{22607}, \country{Germany}}}

\affil[2]{\orgdiv{Physics Department}, \orgname{University of Hamburg}, \orgaddress{\street{Luruper Chaussee 149}, \mbox{\city{Hamburg}, \postcode{22761}, \country{Germany}}}}

\affil[3]{\orgname{Antalya Bilim University}, \orgaddress{\street{07190 Dosemealti}, \city{Antalya}, \country{Turkey}}}

\affil[4]{\orgname{The Hamburg Centre for Ultrafast Imaging}, \orgaddress{\street{Luruper Chaussee 149}, \mbox{\city{Hamburg}, \postcode{22761}, \country{Germany}}}}

\abstract{High-energy (mJ) and high-peak-power (MW) multicycle terahertz (THz) pulses are essential for nonlinear THz spectroscopy and compact accelerator technologies, yet their generation by nonlinear optical frequency conversion remains inefficient and imposes severe demands on femtosecond driving lasers. Amplifying existing THz pulses offers an appealing alternative, but no power-scalable amplifier has been realized in the sub-THz regime. Here, we demonstrate an all-optical THz amplifier operating at 0.35 THz based on the modulation of nanosecond laser pulses by a weak THz field in periodically poled lithium niobate (PPLN). The THz-induced phase modulation is converted into an amplitude modulation using controlled group delay dispersion, forming a tailored pulse train that can efficiently drive high-energy THz generation in a second crystal, thereby amplifying the THz seed. By analogy to electronic klystrons, we term this device the Phlystron, in which the electron beam carrying the power is replaced by a photon beam. In this proof-of-concept experiment, a 3.3-fold increase in THz energy is achieved with commercial crystals. Scaling analysis indicates the potential for higher gain when using large-aperture PPLN devices and multi-stage amplification. The Phlystron thus provides a scalable route to powerful multicycle THz sources driven by readily available narrowband lasers.}


\maketitle

\section{Introduction}\label{sec1}

The lack of compact, high-energy terahertz (THz) sources in the 0.1–1 THz range remains a central obstacle to advancing applications in nonlinear THz spectroscopy \cite{huang_extreme_2024,hebling_highpower_2008,kampfrath_resonant_2013}, strong-field matter interaction \cite{liu_terahertzfieldinduced_2012,han_strong_2023,nicoletti_nonlinear_2016} and THz-driven particle manipulation and acceleration \cite{hibberd_acceleration_2020,curry_meterscale_2018,xu_cascaded_2021,nanni_terahertzdriven_2015,ying_high_2024,zhang_segmented_2018,kealhofer_alloptical_2016}. For compact linear electron accelerators, in particular, multicycle pulses with millijoule-level energies and megawatt peak powers are highly desirable \cite{vahdani_dielectric_2024,wong_compact_2013}. Free-electron lasers can deliver THz radiation approaching this parameter regime \cite{fisher_singlepass_2022,liang_superradiant_2026}, but their size and infrastructure requirements are incompatible with compact accelerator concepts. Laser-driven sources based on optical rectification in nonlinear crystals offer a promising alternative, as the THz radiation directly inherits the peak power of the pump laser \cite{lemery_highly_2020,jolly_spectral_2019,dalton_cryogenically_2024}. However, the intrinsically low optical-to-THz conversion efficiency - typically below one percent - demands femtosecond pump pulses with Joule-level energies to achieve the desired THz pulse energies and corresponding field strengths.

Titanium-sapphire lasers can meet these requirements but are prohibitively expensive at such energies and limited in repetition rate due to the large quantum defect and the associated heat load \cite{moulton_spectroscopic_1986}. By contrast, Yb-based systems offer superior average power scalability \cite{fattahi_thirdgeneration_2014,herkommer_ultrafast_2020,wang_11_2020} but suffer from a narrower gain bandwidth \cite{deloach_evaluation_1993}, often necessitating the use of post-compression techniques. Although spectral broadening via self-phase modulation in hollow-core fibers and multi-pass cells enables pulse shortening by more than an order of magnitude, beam distortions, laser-induced damage and ionization currently prevent operation at Joule-level energies \cite{viotti_multipass_2022,pfaff_nonlinear_2023}. Thus, further scaling broadband drivers alone may not provide a viable path towards THz pulse energies in the millijoule range. A conceptually different strategy is to amplify an existing THz pulse. Various approaches have been explored including amplification in semiconductor quantum cascade structures \cite{jukam_terahertz_2009,kao_amplifiers_2017,oustinov_phase_2010}, graphene-based metamaterials \cite{boubanga-tombet_roomtemperature_2020} or laser-induced plasmas \cite{dai_terahertz_2007} as well as THz parametric amplification in nonlinear crystals \cite{tripathi_terahertz_2014,murate_sixbillionfold_2020}. While each has demonstrated THz gain, constraints in frequency range, achievable gain or extractable pulse energy have so far prevented the realization of a compact power-scalable amplifier for multicycle pulses in the sub-THz regime.

To address this challenge, we introduce a photonic analog of the electronic klystron harnessing the abundant energy of narrowband lasers. In this scheme, which we term the Phlystron, a weak THz seed pulse imprints a phase modulation onto a high-energy nanosecond laser pulse via the electro-optic effect in a periodically poled nonlinear crystal. Dispersive pulse shaping subsequently converts this phase modulation into an amplitude modulation, producing a pulse train that efficiently generates multicycle THz radiation in a second, identical crystal via optical rectification. This process enables coherent energy extraction from the nanosecond pulse, effectively amplifying the THz seed. The sequence of phase modulation, dispersive pulse formation and nonlinear conversion mirrors the electron dynamics in a two-cavity klystron, in which velocity modulation, electron bunching and energy extraction in the catcher cavity occur at a common radio-frequency (RF) resonance \cite{varian_high_1939}. In the photonic analog, the nonlinear crystals with identical phase-matched THz frequency assume the role of the resonant RF cavities, while the modulated nanosecond pulse corresponds to the microbunched electron beam.

\section{Results}\label{sec2}

\begin{figure}[tb]
\centering
\includegraphics[width=\textwidth]{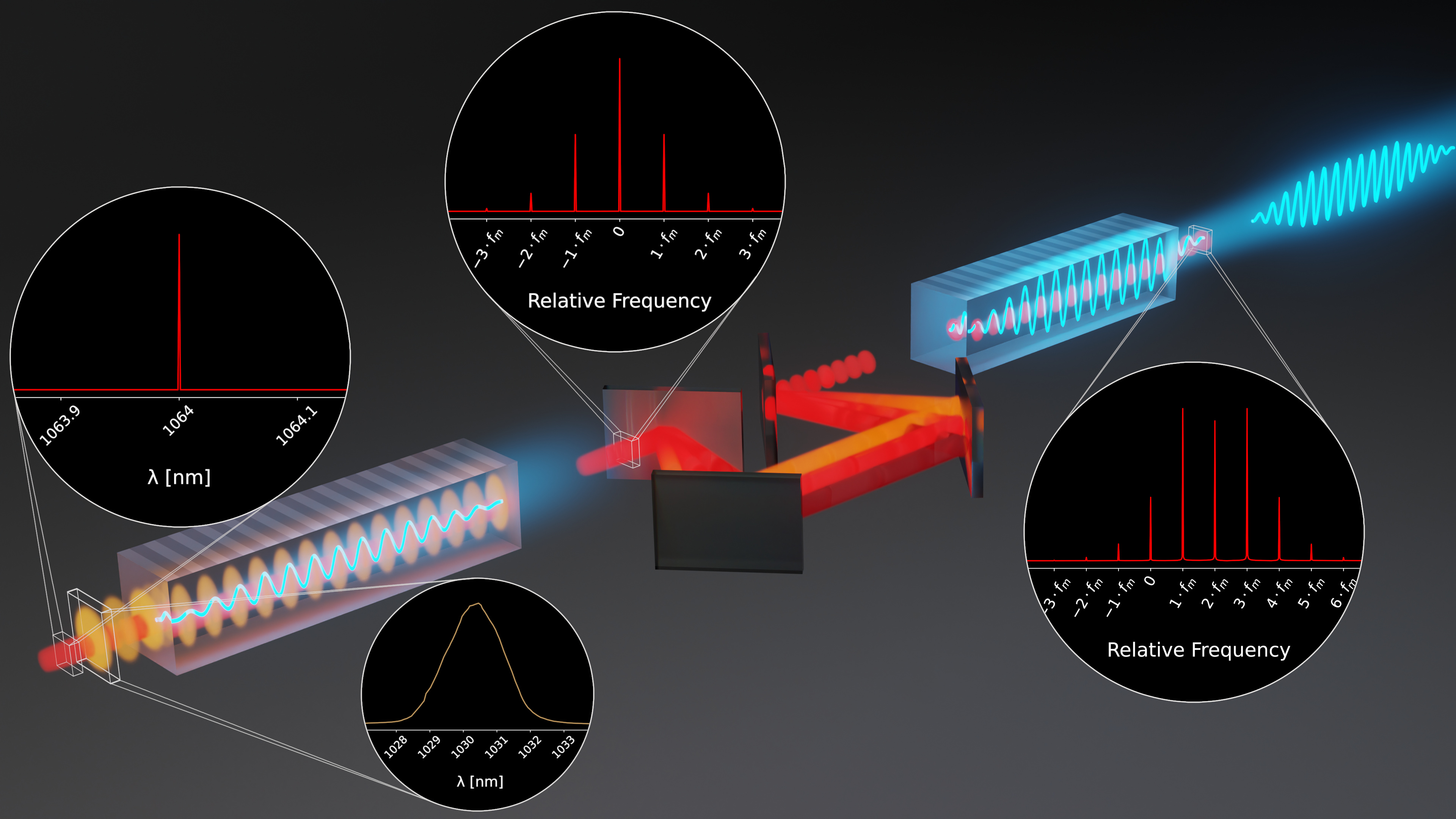}
\caption{\textbf{Schematic of the Phlystron - THz amplifier concept.} In the PPLN modulation crystal (left), the narrowband laser pulse (red) is phase-modulated by a weak multicycle THz pulse (blue), which is generated here with a lower-energetic pulse-train driver (orange). The insets show the respective laser spectrum. After tailoring the spectral phase of the acquired sidebands in a grating compressor, the resulting amplitude modulated pulse is applied to a second PPLN crystal (right) for efficient THz generation producing higher-energetic and effectively amplified THz pulses. The energy extraction from the modulated laser results in a spectral downshift.}
\label{fig1}
\end{figure}

The Phlystron concept was implemented with two cryogenically cooled, commercially available periodically poled lithium niobate (PPLN) crystals (\autoref{fig1}). In the modulation crystal, the 700-ps pulses of a narrowband laser (Nd:YAG, 1064 nm, 10 Hz) were phase-modulated by weak THz pulses, which were generated in situ via optical rectification driven by a synchronized pulse-train laser. However, any externally supplied THz field at the phase-matched THz frequency $f_{THz}$ could be coupled in for amplification. The Yb-based pulse-train laser produces highly regular THz-rate trains of 800 fs pulses with total energies of up to 30 mJ (1030 nm, 10 Hz) \cite{matlis_precise_2023}. Such pulse trains have been shown to enhance the conversion efficiency while mitigating undesired nonlinear effects \cite{matlis_precise_2023,demirbas_advantages_2024}. To ensure temporally uniform phase modulation, the pulse-train length was maximized to 256 pulses such that the THz pulse duration exceeds that of the narrowband pulse (see Methods).

The two laser beams were combined collinearly in the modulation crystal, ensuring spatio-temporal overlap of the narrowband and THz pulses. The close central laser wavelengths allow the crystal poling to simultaneously provide quasi-phase matching for THz generation and THz-induced modulation. Controlled group delay dispersion (GDD) was introduced to the phase-modulated pulses with a grating compressor to tailor the spectral phase of the acquired sidebands, forming a phase-stable pulse train with intraburst repetition rate $f_{THz}$. These secondary pulse trains were then applied to the generation crystal, whose identical specifications ensure that THz generation remains phase-matched at the modulation frequency. Net THz amplification is achieved when the THz output energy exceeds the energy of the modulating THz seed pulse. The amplification factor is hence determined by the optical-to-THz conversion efficiency in the generation crystal and the modulated pulse energy.

\subsection{THz-induced modulation of the narrowband laser}\label{subsec1}

The narrowband pulse energy is limited by the total fluence that can be safely applied to the modulation crystal during long-term operation. The laser-induced damage threshold of PPLN at 78 K was found empirically at a fluence of $\approx 200\,\mathrm{mJ}/\mathrm{cm}^2$ for sub-ps pump pulses at 10 Hz. Maximizing the conversion efficiency $\eta$ in the modulation crystal is therefore crucial to minimize the pulse-train fluence required to generate the modulating THz pulse. To enable a direct comparison of different THz generation methods, we quantify their performance with the slope $\partial\eta /\partial F_0$ of the efficiency curve versus pump fluence $F_0$. Previously, a performance of $1.0\cdot10^{-3}\,\%/(\mathrm{mJ}/\mathrm{cm}^2)$ was demonstrated with trains of 256 pulses and a 2 cm long crystal \cite{matlis_precise_2023}. Longer crystals are expected to be beneficial, as the reduced optical peak intensity in the pulse-train approach mitigates parasitic nonlinear effects that would otherwise accumulate and deplete the pump energy \cite{demirbas_advantages_2024}. Therefore, crystals with lengths of 3 and 5 cm were investigated here, additionally providing an extended interaction length for THz modulation.

\begin{figure}[tb]
\centering
\includegraphics[width=\textwidth]{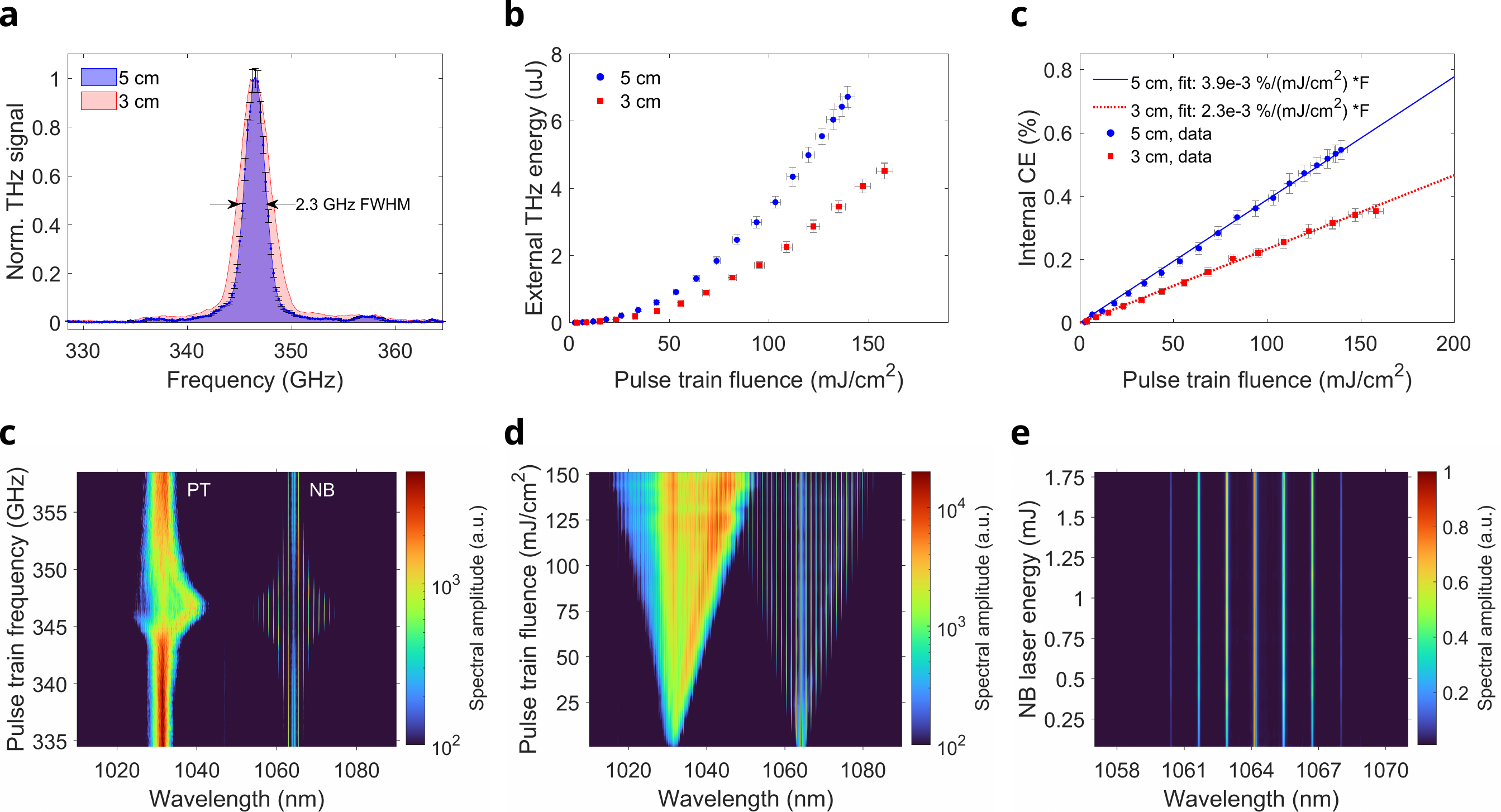}
\caption{\textbf{THz-induced modulation of the narrowband laser.} THz generation with the pulse-train driver in the modulation crystal: \textbf{a} THz tuning curves showing the THz yield versus pulse-train frequency. \textbf{b} and \textbf{c} Corresponding external THz energy at the detection point and internal conversion efficiency (CE) from a 3 and 5 cm long modulation crystal as function of the pulse-train fluence. THz-induced modulation in the 5-cm crystal: \textbf{d} Spectrum of the combined laser system versus pulse-train frequency on the vertical axis (PT - pulse train at 1030 nm, NB - narrowband laser at 1064 nm). This measurement corresponds to the tuning curve scan from (a). \textbf{e} Modulated spectrum versus pulse-train fluence. \textbf{f} Spectrum of the narrowband laser versus its pulse energy for a fixed pulse-train fluence showing the lack of dependence of the modulation process on the energy of the narrowband laser.}
\label{fig2}
\end{figure}

To characterize the modulation process, both the outcoupled THz energy and the combined optical spectrum after the crystal were measured under varying laser parameters. Notably, the THz energy is independent of the presence of the narrowband laser, because the modulation does not consume THz photons. Optimal THz yield requires highly regular pulse trains with an intraburst repetition rate (pulse-train frequency) set to the phase-matched frequency of the crystal to achieve coherent overlap of the generated THz waves. The pulse trains were hence tuned using a tailored process that ensures temporal regularity (see Methods). The phase-matched THz frequency is assessed from the THz tuning curve, which represents the THz yield as function of pulse-train frequency corresponding to the convolution of the PPLN crystal response with the pulse-train excitation spectrum \cite{matlis_precise_2023,demirbas_temperature_2024}. The tuning curves measured with the 3- and 5-cm crystal are both centered at $f_{THz}=346.5\,\mathrm{GHz}$ and symmetric, confirming good tuning, as temporal irregularities would manifest in strong asymmetric side peaks (\autoref{fig2}a). The 5-cm crystal exhibits a remarkably narrow relative tuning curve bandwidth of 0.7 \% (2.3 GHz), emphasizing the narrowband nature of the generated THz pulses. A THz energy of 6.7 $\mu$J was obtained from the 5-cm crystal at 140 mJ/cm$^2$ pump fluence, corresponding to a spectral energy density of $\sim$2.9 mJ/THz and 0.55 \% internal conversion efficiency (\autoref{fig2}b and c). The internal efficiency serves as intrinsic measure of the conversion process decoupled from practical complexities such as THz losses at the exit face or in transport to the detector \cite{matlis_precise_2023} (see Supplementary). The resulting performance of $\partial\eta /\partial F_0=3.9\cdot10^{-3}\,\%/(\mathrm{mJ}/\mathrm{cm}^2)$ doubles the previous efficiency record achieved with a dedicated two-line laser \cite{olgun_highly_2022}. The lower efficiency of the shorter crystal confirms that crystal lengths beyond 3 cm not only provide an extended interaction length but also enhanced conversion when pumping with trains of 256 pulses.

\autoref{fig2}d to f summarize the spectral signatures of the modulation process in the 5-cm crystal. Only when the pulse-train frequency approaches the phase-matched value $f_{THz}$, i.e., when THz radiation is generated, sidebands straddling the 1064 nm line emerge (\autoref{fig2}d). This confirms the THz field as the driver of the modulation rather than direct laser interactions like cross-phase modulation. Due to the highly efficient THz generation, pulse-train fluences lower than 25 mJ/cm$^2$ were sufficient to generate THz pulses that can create intense spectral sidebands (\autoref{fig2}e). Increasing the fluence and thus the THz field strength induces additional sidebands, resulting in a massive broadening of the narrowband laser spectrum of up to 40 nm. Simultaneously, the bandwidth of the pulse-train spectrum also broadens by a similar amount, arising from back-action of the THz field. The spectral center of mass shifts from 1030 nm towards longer wavelengths, as is characteristic of cascaded THz generation \cite{lemery_highly_2020,demirbas_advantages_2024,olgun_highly_2022,mosley_largearea_2023}. The degree of spectral broadening of the narrowband pulses is independent of their energy, as the phase modulation originates from a periodic refractive-index modulation of the crystal induced by the THz field (\autoref{fig2}f). In the photon picture, the symmetric sideband structure reflects balanced sum- and difference-frequency generation processes that conserve the total THz photon number. The measured spacing between adjacent sidebands $\Delta\lambda=1.27\,\mathrm{nm}$ matches the THz frequency well.

\subsection{Conversion of phase to amplitude modulation}\label{subsec2}

A direct spectral phase characterization of the modulated pulses is impractical due to their long duration and complex spectral structure. Instead, we rely on the theoretical model of the traveling-wave modulator (see Supplementary). Assuming a continuous-wave THz field, the narrowband pulses acquire a sinusoidal phase shift oscillating at the modulation frequency $f_m=f_{THz}$ with modulation depth $\theta =-\frac{2}{\pi}\frac{\omega_0 n_e^3}{2c}r_{33}E_{THz}\cdot L$. Here, $L$ is the crystal length, $n_e$ the extraordinary refractive index of lithium niobate, $r_{33}$ its electro-optic coefficient and $E_{THz}$ the THz electric field amplitude. The resultant electric field of the narrowband pulses thus takes the form $E_{op}(t,L)\propto\sqrt{I_0}e^{-i\theta\sin{(f_m t)}}$. The sinusoidal phase modulation corresponds to a frequency modulation, which is periodic in time and has a maximum chirp $\left( \frac{\partial\omega}{\partial t} \right)_{max}=\pm (2\pi f_m)^2\theta$. By applying a GDD to compensate the chirp in one half of the period, the phase modulation is converted into a periodic temporal intensity modulation. For arbitrary $f_m$, this compression can be quantified using a dimensionless bunching parameter $B(\theta)=-\left( \frac{\partial\omega}{\partial t} \right)\cdot\left. \frac{\partial^2\phi}{\partial\omega^2} \right|_{\omega_0}$ with the spectral phase $\phi(\omega)$ introduced by the dispersive element \cite{kobayashi_optical_1988}. Optimal compression is typically reached with $B_{opt}(\theta)\approx 1$.

\begin{figure}[b]
\centering
\includegraphics[width=\textwidth]{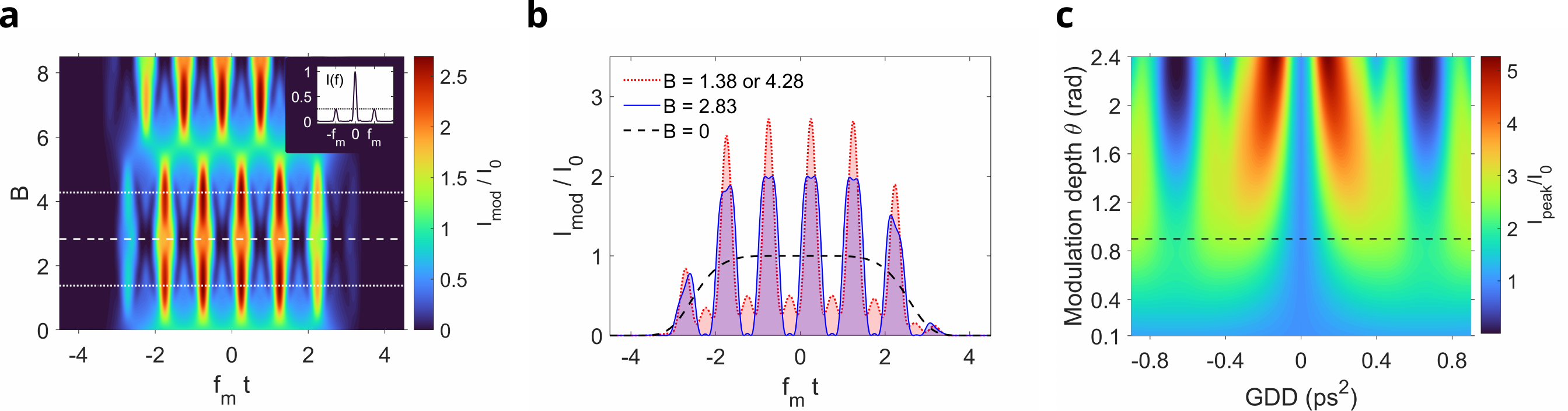}
\caption{\textbf{Conversion of phase to amplitude modulation.} Numerical calculation of the pulse formation by passing a phase-modulated super-Gaussian pulse of order 3 with an FWHM duration of $5f_m t$ through a dispersive element: \textbf{a} Modulated intensity $I_{mod}$  versus normalized time and bunching parameter $B$ for modulation depth $\theta=0.9$. The data is normalized to the peak intensity of the initial pulse $I_0$. The inset shows the corresponding modulated spectrum. The three white lines indicate the $B$-values for which the intensity reaches the global maximum (dotted) and minimum (dashed). \textbf{b} Modulated intensity profiles along the indicators in (a) with the unmodulated pulse ($B=0$) as reference. \textbf{c} Peak intensity of the modulated pulse $I_{peak}$ at time $f_m t=\pm 1/4$ (at which the sub-pulses form for pos./neg. GDD) versus $\theta$ and GDD from the dispersive element assuming $f_m=346.5\,\mathrm{GHz}$. The dashed line indicates $\theta=0.9$.}
\label{fig3}
\end{figure}

For operation of the THz amplifier, low modulation depths are preferred, as this allows the available pulse-train energy to be distributed over a larger area, which in turn supports higher narrowband pulse energies and thus greater THz output. Notably, a shallow modulation of $\theta\approx 0.9$ yields first-order sidebands with 25 \% relative intensity, which, with appropriate dispersion, is sufficient to fully modulate the temporal intensity (\autoref{fig3}a). In this specific case, the model predicts distinct pulse formation reaching maximum peak intensity at $B=1.38$ and $B=4.28$. However, the corresponding intensity profiles exhibit additional spikes between the sub-pulses (\autoref{fig3}b). At the intermediate value of $B=2.83$, the intensity periodically vanishes, creating well-separated sub-pulses. With increasingly large $B$, the modulation reverts to a pure phase modulation. For $f_m=346.5\,\mathrm{GHz}$, the considered values of $B=1.38$, 2.83 and 4.28 correspond to 0.32, 0.66 and 1.01 ps$^2$ GDD, respectively. Increasing the modulation depth reduces the GDD required to maximize the intensity modulation contrast, while enabling shorter sub-pulses and thus higher peak intensity due to the broader spectrum (\autoref{fig3}c). As is characteristic of sinusoidal phase modulation, both positive and negative dispersion lead to pulse formation. At the highest $\theta$ shown in \autoref{fig3}c, optimal compression is achieved in a narrow range around $\pm 0.2\,\mathrm{ps}^2$, whereas shallow modulation ($\theta <1$) offers a broader GDD window around $\pm 0.66\,\mathrm{ps}^2$ at the cost of peak intensity.

\subsection{THz generation with amplitude modulated pulses and THz amplification}\label{subsec3}

\begin{figure}[b]
\centering
\includegraphics[width=\textwidth]{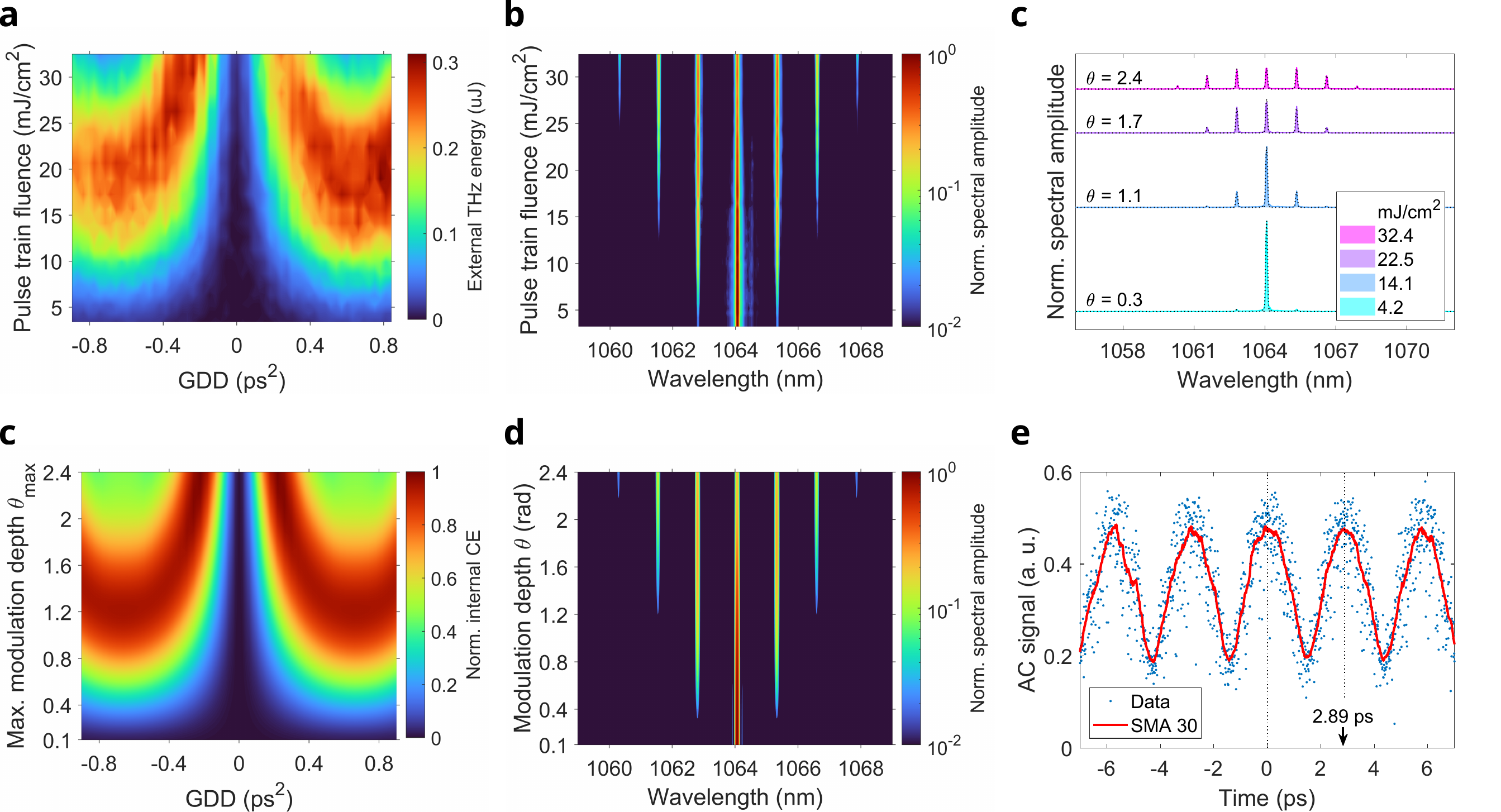}
\caption{\textbf{THz generation with amplitude modulated pulses.} \textbf{a} External THz energy generated with the modulated laser at fixed fluence versus applied GDD and pulse-train fluence used in the modulation stage.
\textbf{b} Narrowband laser spectrum normalized to the unmodulated 1064 nm line for the pulse-train fluences from (a) (the scale of the vertical axis is the same). \textbf{c} Modulated spectrum from (b) for selected pulse-train fluences with the matching calculated spectrum including 22 \% unmodulated intensity (black dotted lines, see (e)). \textbf{d} Numerical simulation of the conversion efficiency as function of the applied GDD and the maximum modulation depth $\theta_{max}$ in the beam after accounting for the spatial dependence of the modulation depth. \textbf{e} Calculated modulated spectrum normalized to the unmodulated 1064 nm line for the $\theta$-values fitting best to the measured spectra from (b). \textbf{f} Autocorrelation trace (AC) of the amplitude modulated pulses ($F_{train}\approx 17\,\mathrm{mJ}/\mathrm{cm}^2$). The raw data is shown together with a simple moving average over 30 data points (SMA 30).}
\label{fig4}
\end{figure}

Monitoring the THz generation in the second PPLN crystal provides an indirect yet robust measure of the amplitude modulation while bypassing the need for extensive time-domain characterization. A Martinez stretcher providing up to $\pm 0.9\,\mathrm{ps}^2$ of GDD was employed to demonstrate compression across both dispersion regimes (see Methods) \cite{martinez_3000_1987}. As expected, THz emission is absent at zero GDD, but emerges with increasing GDD magnitude as the phase modulation is converted into an amplitude modulation (\autoref{fig4}a). The GDD required to reach maximum THz energy decreases with higher pulse-train fluence and modulation depth. The corresponding modulated spectra are shown in \autoref{fig4}b. While the THz yield generally follows the calculated peak-intensity map (\autoref{fig3}c), a broad plateau of high yield persists at intermediate GDD and pulse-train fluence around 0.66 ps$^2$ and 20 mJ/cm$^2$, respectively. This plateau originates from the specific temporal profile of the modulated intensity. Intensity spikes in between the sub-pulses excite out-of-phase THz waves that cause destructive interference and hence reduce the THz yield (see \autoref{fig3}b). Cleaner temporal profiles with well-separated pulses, however, avoid this effect and compensate for the lower peak intensities, maintaining high efficiency even at intermediate modulation depth.

Numerical simulations of the THz generation support this interpretation (see Methods). Due to the spatial intensity variation of the Gaussian THz beam, the modulation depth and hence the conversion efficiency vary across the beam profile. The modulated spectra were measured by sampling the beam in the center with an optical fiber, such that they reflect the maximally achieved spectral broadening (see Supplementary). This maximum modulation depth $\theta_{max}$ was extracted by fitting the data from \autoref{fig4}b to calculated spectra resulting in $\theta_{max}=2.4$ for the highest pulse-train fluence (\autoref{fig4}c and e). While the carrier line at 1064 nm should ideally vanish at this modulation depth, approximately 22 \% of the intensity remains unmodulated. This is attributed to imperfect spatial overlap between the THz and the narrowband laser beam, alongside a residual post-pulse, which is indistinguishable in spectral measurements (see Supplementary). After accounting for the spatial distribution of the modulation depth, the simulated efficiency (\autoref{fig4}d) closely reproduces the trends observed in the measured THz energy from \autoref{fig4}a. Independent confirmation of the amplitude modulation is provided by intensity autocorrelation (\autoref{fig4}f). The autocorrelation peaks are separated by 2.89 ps, corresponding to a 346.1 GHz repetition rate, while the finite signal between peaks indicates a sinusoidal intensity modulation rather than isolated sub-pulses, consistent with operation at low modulation depth.

\begin{figure}[b]
\centering
\includegraphics[width=\textwidth]{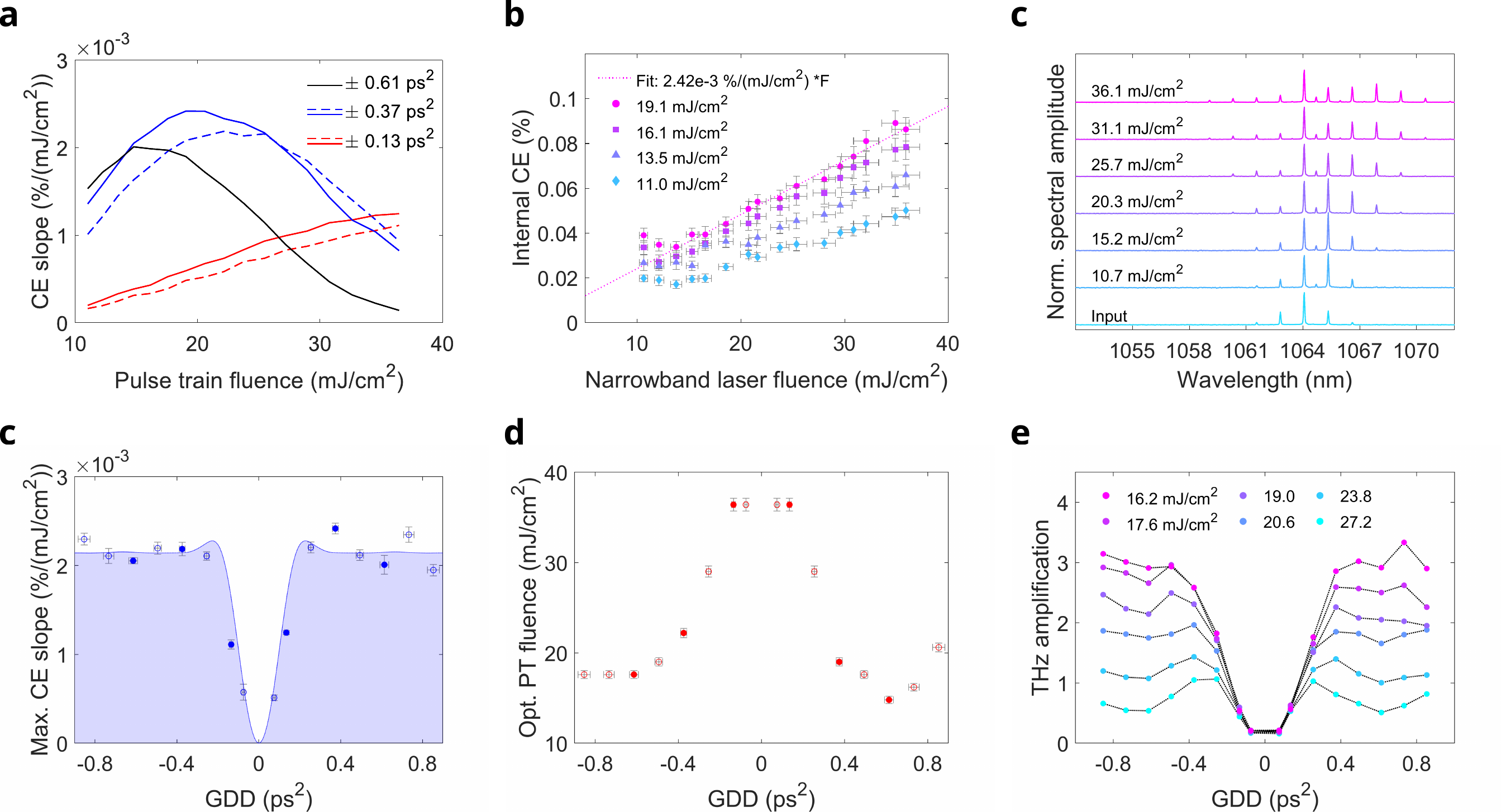}
\caption{\textbf{Characterization of the THz amplification.} \textbf{a} THz generation performance (conversion efficiency (CE) slope) of the modulated pulses as function of the pulse-train fluence used for the modulation for selected negative (dashed lines) and positive (solid lines) GDD values. \textbf{b} Internal CE versus narrowband laser fluence for $+0.37\,\mathrm{ps}^2$ applied GDD and different pulse-train fluences. The line plot is a linear fit to the data with highest efficiency ($F_{train}=19.1\,\mathrm{mJ}/\mathrm{cm}^2$). \textbf{c} Cascaded spectrum of the modulated laser after THz generation for selected pump fluences under optimum modulation conditions from (b). \textbf{d} Maximum achieved performance as a function of the applied GDD. Filled circles indicate the values for which the data is shown in (a). The dotted line is a guide to the eye according to the simulation data in \autoref{fig4}d. \textbf{e} Corresponding optimum pulse-train fluence versus GDD. \textbf{f} THz energy amplification factor as function of GDD for selected pulse-train fluences.}
\label{fig5}
\end{figure}

To quantify the effectiveness of the modulated pulses, the generated THz energy was measured as function of the narrowband laser fluence applied to the generation crystal for several stretcher settings. Higher GDD magnitudes reduce the pulse-train fluence required for maximum THz generation performance, whereas small GDD ($\approx\pm 0.13\,\mathrm{ps}^2$) results in minimal THz yield due to the absence of pronounced amplitude modulation (\autoref{fig5}a). As expected, the performance for positive and negative GDD is nearly identical. \autoref{fig5}b shows the internal conversion efficiency for the case of $+0.37\,\mathrm{ps}^2$ GDD, which provided the highest performance of $2.42\cdot10^{-3}\,\%/(\mathrm{mJ}/\mathrm{cm}^2)$ with 19.1 mJ/cm$^2$ pulse-train fluence. The fluence at the generation crystal was limited to $\approx 40\,\mathrm{mJ}/\mathrm{cm}^2$ by the fluence constraints in the modulation stage and optical losses in the compressor and beam-shaping optics. Efficient THz generation is corroborated by the spectral downshift of the modulated laser (\autoref{fig5}c). Cascaded frequency conversion manifests as distinct spectral lines on the long-wavelength side that intensify with increasing pump fluence, in contrast to the continuous shift of the pulse-train laser spectrum (\autoref{fig2}e). This reflects the well-defined periodicity of the modulated pulses with correspondingly narrow spectral lines, whereas the timing of the pulses in the initial pulse train is imperfect with respect to the optical cycle, which results in a continuous spectrum.

The THz generation performance of the modulated pulses is summarized in \autoref{fig5}d. At sufficiently high GDD magnitude, the maximum performance saturates at $\partial\eta /\partial F_0=2.2\cdot10^{-3}\,\%/(\mathrm{mJ}/\mathrm{cm}^2)$, approximately half the value achieved with the pulse-train driver in the modulation crystal (see \autoref{fig2}c). The corresponding optimal pulse-train fluence is $\approx 17\,\mathrm{mJ}/\mathrm{cm}^2$ (\autoref{fig5}e). The lower performance is partly attributed to stress-induced beam distortions introduced in the modulation crystal, which can be mitigated by optimized stress-free crystal mounting (see Supplementary). Nonetheless, effective THz amplification is achieved because the energy of the modulated pulse significantly exceeds that of the initial pulse train. The amplification factor is defined as the ratio of the maximum THz energy attained from the generation crystal to the THz seed pulse energy. It decreases with increasing pulse-train fluence, as deeper modulation requires more energetic seed pulses (\autoref{fig5}f). Consequently, the maximum amplification factor of 3.3 (5.2 dB) is obtained at shallow modulation (16–18 mJ/cm$^2$ pulse-train fluence), corresponding to an increase in THz energy from 90 nJ to 0.3 $\mu$J.

\section{Discussion}\label{sec4}

While the presented experiments were conducted with small-aperture crystals and mJ-level pump energies as a proof of concept, the scheme is intrinsically scalable to higher energies. At the optimal pulse-train fluence of 17 mJ/cm$^2$, the narrowband laser fluence can be 11 times higher without exceeding the sustainable total fluence of 200 mJ/cm$^2$. Due to the different Gaussian beam sizes (see Methods), this corresponds to an energy ratio reduced by a factor of two (5.5), and hence a comparable upper limit for the achievable amplification factor when assuming the same conversion efficiency in both crystals. The Phlystron effectively transforms the continuous pulse-train spectrum into a series of narrowband lines, thereby imposing a well-defined spectral structure on the optical driver that is ideally suited for optical-to-THz conversion \cite{ravi_raman_2020}. In the time domain, the resulting pulse trains are perfectly regular suggesting that THz generation in the second crystal could be even more efficient than in the first. With ideal flat-top beam profiles, an 11-fold THz amplification becomes feasible, because all beams would have equal size, such that the peak fluence and energy ratio are equal. For the full 30 mJ pulse-train energy, this corresponds to modulation of a 330 mJ narrowband pulse.

Safe operation at such energies requires scaling of the crystal apertures. Large-aperture PPLN devices are realized either by poling and stacking bulk crystals with up to 10 mm thickness or by assembling stacks of manually oriented wafers \cite{lemery_highly_2020,dalton_cryogenically_2024,mosley_largearea_2023,matlis_scaling_2024}. Both approaches pose engineering challenges in maintaining THz-wave coherence and minimizing loss due to gaps and crystal interfaces, respectively. With flat-top beams of 15 mm diameter, the total fluence remains already below the damage threshold, allowing internal THz pulse energies of up to $330\,\mathrm{mJ}\cdot 3.9\cdot 10^{-3}\,\%/(\mathrm{mJ}/\mathrm{cm}^2)\cdot 200\,\mathrm{mJ}/\mathrm{cm}^2 = 2.6\,\mathrm{mJ}$ based on the demonstrated conversion efficiencies (\autoref{fig2}c). THz amplification thus provides a new pathway towards mJ-level multicycle THz sources in the sub-THz regime while eliminating the need for large-scale high-energy femtosecond laser drivers.

Increasing the energy of the pulse-train driver or adopting a multi-stage architecture as in laser amplifier chains can enable the modulation of even more energetic pulses. Each additional modulation stage would yield a THz amplification by one order-of-magnitude, given that the crystal apertures are scaled appropriately (see Supplementary). While demonstrated at 347 GHz, operation at higher frequencies is possible by simply reducing the PPLN poling period. Because the THz generation efficiency increases with frequency, the amplification should even improve at higher THz frequencies until the increasing THz absorption cancels this advantage \cite{carletti_nonlinear_2023,vodopyanov_optical_2006}. Hence, efficient operation is anticipated up to $\sim$1 THz when using lithium niobate crystals.

Beyond THz amplification, the modulation stage alone acts as a THz frequency traveling-wave modulator, significantly exceeding the bandwidth limit of conventional RF devices near 100 GHz. While single-cycle THz transients have previously enabled supercontinuum generation \cite{giorgianni_supercontinuum_2019,vicario_subcycle_2017}, cross-phase modulation \cite{shen_nonlinear_2007,koulouklidis_observation_2020} and electro-optic time lensing \cite{shen_electrooptic_2010}, electro-optic modulation (EOM) driven by laser-generated multicycle THz fields had not been demonstrated yet. THz-EOM offers a practical and cost-efficient pathway to creating highly energetic trains of ultrashort pulses with THz intraburst repetition rate and comb-like spectra, and thus may open up new possibilities for plasma-wave excitation \cite{umstadter_nonlinear_1994,cowley_excitation_2017} and laser material processing \cite{kerse_ablationcooled_2016,gaudiuso_laser_2021,wang_ultrafast_2020,bonamis_systematic_2020}, where GHz-to-THz burst-mode operation improves ablation efficiency.

\section*{Methods}

\textbf{Laser system.} The narrowband laser is a commercial Q-switched Nd:YAG system (Amplitude) operating at 1064 nm and 10 Hz with pulse energies up to 2.2 J. Here, it was operated with a home-built fiber seed laser based on ultrafast pulse chopping to reduce the FWHM pulse duration to 698 ps \cite{liu_10hz_2021}. The Super-Gaussian output beam was spatially filtered and resized resulting in a Gaussian beam with $2w_{NB}\approx 2.5\,\mathrm{mm}$ 1/e$^2$-intensity diameter in the far-field to optimally fill the modulation crystal aperture. To allow a direct comparison of the THz yield obtained with the modulated laser and the pulse-train driver, the beam size at the generation crystal was set to match that of the pulse-train laser. The beam distortions introduced by the modulation crystal were compensated with a set of cylindrical telescopes to recover a decent beam shape.

The pulse-train laser is a home-built Yb-based chirped pulse amplification system providing regular trains of ultrashort pulses at 1030 nm and 10 Hz repetition rate with up to 30 mJ pulse train energy \cite{matlis_precise_2023}. A motorized pulse division setup comprising eight polarization-based interferometers splits pulses from a 40 MHz fiber front-end into trains of to $2^n$ pulses with tunable pulse-train frequency, when $n$ interferometers are active. Following two-stage amplification, all pulses are compressed to a transform-limited FWHM duration of about 800 fs. The pulse-train regularity was optimized using intensity autocorrelations measured with a commercial device (APE GmbH). Decisive individual pulse delays were fine-tuned via spectral interference with a reference pulse of well-known delay.

Because the pulse train-frequency must be precisely matched to the THz frequency $f_{THz}$, the chosen pulse number of $N_{train}=256$ corresponds to a temporal pulse-train length of $(N_{train}-1)/f_{THz}=737\,\mathrm{ps}$. To achieve a spatially homogenous modulation of the narrowband pulse, the THz beam is ideally flat-top. However, since the pulse-train laser has a Gaussian beam profile, the generated THz beam is Gaussian as well and reduced in size by the second-order nonlinear conversion. The pulse-train laser size was hence set to $2w_{PT}=\sqrt{2}\cdot 2w_{NB}\approx 3.5\,\mathrm{mm}$ to match the THz beam size to that of the narrowband laser.

\medskip\noindent
\textbf{Laser synchronization and beam combination.} The two lasers were synchronized electronically using a digital delay generator (SRS DG645) referenced to the fiber oscillator of the pulse-train laser, enabling relative timing control with 1 ps. Proper timing of pulse train and narrowband pulse was found via non-collinear sum frequency generation in a type I BBO crystal. The temporal overlap of narrowband and modulating THz pulse was subsequently refined by maximizing the energy depletion of the 1064 nm laser line. The laser beams were overlapped spatially with a broadband 50:50 beam combiner and ensured to be perfectly collinear using a CCD camera (DataRay Inc.) in the near and far field. The polarizations were adjusted in front of the combiner. 

\medskip\noindent
\textbf{THz modulation and generation.} Two uncoated 5\%-MgO-doped PPLN crystals (HC Photonics) with $4\times 4\,\mathrm{mm}^2$ aperture, 50 mm length, and 330 $\mu$m poling period were used in the final setup. Lithium niobate was chosen as nonlinear medium due to its exceptionally high nonlinearity and resistance to laser-induced damage. Both crystals were cooled to 78 K in liquid-nitrogen cryostats to reduce THz absorption. The phase-matched frequencies were determined from tuning scans as 346.5 GHz and 345.8 GHz, respectively. The difference of only 0.7 GHz enables close to resonant excitation of the generation crystal with the modulated pulses. All fields were polarized along the optical z-axis, as the crystals were poled in this direction to exploit the highest nonlinear and electro-optic coefficient $d_{33}$ and $r_{33}=32\,\mathrm{pm/V}$, respectively, leading to an effective bulk susceptibility for THz generation of $\chi_0^{(2)}=336\,\mathrm{pm/V}$ \cite{ravi_pulse_2016}. While the maximum number of THz field cycles generated by a pulse train in a PPLN crystal with $N_{PPLN}$ poling periods is given by $N_{train}+N_{PPLN}-1$, the actual number may be reduced by THz absorption, which also gives the THz pulse envelope its characteristic shape \cite{ravi_pulse_2016}. $N_{train}=256$ and $N_{PPLN}=50\,\mathrm{mm}/0.33\,\mathrm{mm}\approx 152$ results in a nominal THz pulse duration obtained from the modulation crystal of $407/f_{THz}\approx 1.175\,\mathrm{ns}$.

\medskip\noindent
\textbf{THz energy and spectral measurements.} In both stages, the THz radiation was collimated and focused using 2-inch off-axis parabolic mirrors and separated from the laser beams with a single 10 mm thick Teflon plate (see Supplementary for THz transmittance data). THz pulse energies were measured in the focus with a pyroelectric detec-tor (Gentec SDX-1152) cross-calibrated to a THz sensor, which had been calibrated by the Physikalisch-Technische Bundesanstalt (PTB). Infrared spectra were recorded using a high-resolution fiber-coupled spectrometer (ASEQ Instruments). A single-mode fiber with 5.3 $\mu$m mode field diameter was used to probe the beams.

\medskip\noindent
\textbf{Dispersive pulse shaping.} GDD was introduced to the phase-modulated pulses using two identical transmission gratings with 1000 lines/mm operated at the Littrow angle (32$^{\circ}$) in a Martinez stretcher configuration \cite{martinez_3000_1987}. The first grating was imaged with a non-magnifying telescope while the second grating was mounted on a motorized translation stage with a travel range of $\pm 150\,\mathrm{mm}$ around the image position, enabling continuous tuning of the GDD from of -0.90 ps$^2$ to +0.90 ps$^2$. A sketch of the complete experimental setup is provided in the Supplementary.

\medskip\noindent
\textbf{THz generation simulations.} The numerical simulations of THz generation with amplitude-modulated pulses were based on a one-dimensional model including sum- and difference-frequency generation (and thus pump depletion and cascading), self-phase modulation and THz absorption \cite{ravi_pulse_2016}. \autoref{fig4}d is based on 2209 simulation runs using temporal intensity profiles derived from 47 modulation depths and 47 GDD values. To reduce computation time, a crystal length of only 5 mm was assumed. Therefore, only relative efficiencies normalized to the maximum value are shown.

\bigskip
\bibliography{PhlystronPaper}

\backmatter

\section*{Data availability}

Data underlying the results presented in this paper are not publicly available at this time but may be obtained from the corresponding author upon reasonable request.

\section*{Code availability}

The relevant computer codes supporting this study are available from the corresponding author upon reasonable request.

\section*{Acknowledgements}

This work was supported by the Deutsches Elektronen-Synchrotron (POV-IV - Matter MML-DMC), the European Research Council under the European Union’s Seventh Framework Programme (FP7/2007-2013) through Synergy Grant AXSIS (609920), and the Deutsche Forschungsgemeinschaft (Project No. 405983224). C. Rentschler acknowledges the Max Planck School of Photonics. The authors thank Marvin Edelmann, Martin Kellert and Jelto Thesinga for technical support with the laser systems.

\section*{Author contributions}

C.R. set up and performed the experiments, carried out the data analysis and simulations, and wrote the manuscript. U.D. and Z.Z. built and developed the pulse-train laser system based on a concept by N.H.M. and M.P.; J.N. contributed the 3D-rendered schematic and assisted in measurements. K.R. developed the initial simulation code, which was further extended and used by C.R.; M.P. led the design and implementation of the laser systems. N.H.M. and F.X.K. conceived and supervised the research. All authors reviewed the manuscript.

\section*{Competing interests}

The authors declare no competing interests. 

\section*{Additional information}

\bmhead{Supplementary information}
The online version contains supplementary material.

\bmhead{Correspondence}
Correspondence and requests for materials should be addressed to Franz Kärtner.


\end{document}